\documentstyle[12pt]{article}

\begin{document}

\hfill \vbox{\hbox{UCLA/95/TEP/14}
             \hbox{hep-th/9505005} }

\begin{center}

{\Large \bf Topology on the lattice; 2d Yang-Mills theories with a theta term
}  \\[10mm]

{\bf Tam\'as G.\ Kov\'acs}\footnote{e-mail: kovacs@physics.ucla.edu}
and {\bf E.T.\ Tomboulis}\footnote{e-mail: tomboulis@physics.ucla.edu} \\
{\em Department of Physics UCLA,
Los Angeles, California 90024-1547, U.S.A.} \\[5mm]

{\bf Zsolt Schram}\footnote{e-mail: schram@tigris.klte.hu} \\
{\em Department of Theoretical Physics,
Kossuth Lajos University,
H-4010 Debrecen, POBox 5, Hungary} \\

\end{center}

\section*{Abstract}

We study two-dimensional U($N$) and SU($N$) gauge theories with a topological
term on arbitrary surfaces. Starting from a lattice formulation we derive
the continuum limit of the action which turns out to be a generalisation of
the heat kernel in the presence of a topological term. In the continuum limit
we can reconstruct the topological information encoded in the theta term.
In the topologically trivial cases the theta term gives only a trivial
shift to the ground state energy but in the topologically nontrivial ones
it remains to be coupled to the dynamics in the continuum.
In particular for the U($N$) gauge
group on orientable surfaces it gives rise to a phase transition at $\theta=
\pi$, similar to the ones observed in other models. Using the equivalence
of 2d QCD and a 1d fermion gas on a circle we rewrite our result in the
fermionic language and show that the theta term can be also interpreted as
an external magnetic field imposed on the fermions.

\vfill
\pagebreak

\section{Introduction}

Recently there has been a renewed interest in studying lattice models with
a topological term. This was motivated mostly by the hope that the
long-standing strong CP problem can be solved \cite{Schierholz}. Recall that
in four dimensional QCD one can in principle add a CP violating term,
the so called theta term to the conventional Yang-Mills action. In nature
however, strong interactions seem to respect the CP symmetry, so the coupling
constant of this term $(\theta)$ must be zero or at least very small. The
question is why nature is forced to choose the special value $\theta=0$.
Up to now Monte Carlo studies of this model have been very limited mostly
due to the following two difficulties. At first the theta term in Euclidean
space-time is purely imaginary which makes the Boltzmann weight of some
states negative. It means that the Boltzmann weight cannot be interpreted
as a probability measure. This difficulty, however, has been largely overcome
by some new simulation techniques developed in the last five years
\cite{Wiese}. The other problem arising in Monte Carlo simulations is related
to the fact that the theta term is of topological nature. In four
dimensional QCD on the sphere the theta term is proportional to the instanton
number, an integer that does not change upon continuous deformations of the
field configuration. On the lattice however, the notion of continuity is
completely lost, or in other words continuity does not impose any restriction
on the deformation of the degrees of freedom at different locations. As a
result there are no disconnected instanton sectors and the very definition
of the topological charge is ambiguous. A possible way of defining the
topological charge is by picking a smooth
``interpolating'' field that is in some sense close to the given lattice
configuration and assigning the topological charge of the interpolating field
to the lattice configuration. This procedure is highly nontrivial and it
is not clear what happens in the quantum continuum limit. An interesting
question one can ask is whether the topological information encoded in the
theta term is retained in the continuum limit.

In this paper we shall study this question in two-dimensional
U($N$) and SU($N$) Yang-Mills theories defined on surfaces of different
topologies. In this case the topological term is proportional to the
first Chern number of the underlying bundle. Our choice of the lattice
theta term is the most naive, the only requirement it satisfies is that
in the naive continuum limit it gives the Chern number. For U($N$) bundles
on orientable surfaces the first Chern number can take any
integer value depending on the ``twist'' in the bundle.
On the other hand it is identically zero for
the SU($N$) groups and also for U($N$) on non-orientable surfaces. Even in
the latter cases, the same ``theta-term'' as for the nontrivial U($N$) case,
can be included in the lattice action. With our choice of the lattice
theta term it does not vanish for any of the above discussed cases,
it does not necessarily give an integer topological charge on the lattice,
but it has the correct continuum behaviour. We shall show that the topological
information encoded in the theta term can be completely recovered in the
continuum limit. This means that in those cases where the Chern number is
trivial, the theta term completely decouples from the dynamics, giving only
a constant shift of the free energy. On the other hand, for U($N$) gauge
groups on orientable surfaces the theta term survives the continuum limit and
it gives rise to a phase transition at
$\theta = \pi$, similar to the ones observed in the 2d U(1)
gauge model \cite{Wheater} and in CP($N$) models at
strong coupling \cite{Seiberg}.

The plan of the paper is as follows. In section 2 we calculate the partition
function of 2d U($N$) and SU($N$) lattice Yang-Mills theories with a theta term
using the standard group character (``strong coupling") expansion. In section
3 we derive the continuum limit of the action which will turn out to be
the heat kernel with an additional contribution from the theta term. We
interpret the continuum partition function in terms of the underlying topology
and show the presence of a phase transition at $\theta = \pi$ in the
topologically nontrivial cases. Recently it has been shown that 2d QCD is
equivalent to a one dimensional gas of fermions \cite{Minahan}. In section
4 we discuss our results in this language. In particular we show that
the topological term can be interpreted as an external
magnetic field imposed on this fermion gas. Finally in section 5 we
summarise our results and discuss
their possible relevance to more realistic models with a topological term.

\section{The partition function on the lattice}

In this section we derive an expression for the partition function of 2d
Yang-Mills theories with a theta term. We use the standard strong coupling
character expansion. The expression of the partition function in terms of
the character expansion coefficients is well known on arbitrary surfaces, see
e.g. \cite{Rusakov, Witten}, so we shall just quote this result.
We then calculate explicitly the character
expansion coefficients for the specific lattice action that we pick.

We use the conventional Wilson lattice gauge model; the degrees of freedom
are gauge group elements associated to links of a two dimensional square
lattice. The gauge group $G$ can be thought of as either SU($N$) or U($N$)
until some specific choice is stated. The action for a configuration
$[U_l]$  is given by
\begin{equation}
  {\cal S} [U_l] = \sum_p  s(U_p),
\end{equation}
where $U_p$ is the product of link variables around the plaquette $p$,
$s$ is a conjugacy class function on $G$ (i.e.\ $s(g^{-1} U g = s(U),
\forall g,U \in G$) and the summation is for all the plaquettes. The
partition function of a finite lattice consisting of $A$ plaquettes is
\begin{equation}
 Z_A(g) = \sum_{[U_l]} \exp \left( -\frac{N}{g^2} \sum_p s(U_p) \right),
\end{equation}
where $g$ is the gauge coupling and the summation is for all the
configurations.

Since the Boltzmann factor on each plaquette is a conjugacy class function,
it can be expanded in irreducible characters of $G$ as
\begin{equation}
 \exp \left( -\frac{N}{g^2} s(U) \right) = \sum_r d_r \Lambda_r(N/g^2)
 \chi_r(U),
      \label{eq:charexp}
\end{equation}
where
\begin{equation}
 \Lambda_r(N/g^2) = \frac{1}{d_r} \int_G dU \exp \left( -\frac{N}{g^2}
 s(U) \right) \, \chi^*_r(U).
     \label{eq:charcoeff}
\end{equation}
Here $r$ labels the irreducible representations of $G$, $d_r$ is the dimension
and $\chi_r$ is the character of $r$. Substituting (\ref{eq:charexp}) into
the expression of the partition function, the link variables can be
integrated out using the orthogonality of the characters. For a closed
orientable surface of genus $g$ we obtain \cite{Rusakov}
\begin{equation}
 Z_A(g) = \sum_r d_r^{2-2g} \Lambda_r^A (N/g^2)
     \label{eq:pf}
\end{equation}
For non-orientable surfaces the summation is restricted to
self-conjugate representations and also there is an additional constant
factor \cite{Witten}, which will be unimportant here, so we omit it.

Up to this point we have not specified the form of the action, (\ref{eq:pf})
holds for any gauge invariant action. The information specific to the
action is entirely contained in the $\Lambda_r$'s. We now choose for the
Yang-Mills part the usual Wilson action
\begin{equation}
 s_w(U) = \frac{1}{2} ( \mbox{tr} U + \mbox{tr} U^\dagger )
\end{equation}
To make the analytic calculations feasible we choose
\begin{equation}
 s_\theta (U) = \frac{1}{2} ( \mbox{tr} U - \mbox{tr} U^\dagger)
\end{equation}
for the theta term. Obviously this have the desired continuum behaviour
which can be easily seen in exactly the same way as the continuum limit
of the Wilson term. While in the continuum limit $s_w(U)$
is proportional to the square of the field strength (up to a trivial
additive constant),  the term $s_\theta (U)$ goes linearly with the field
strength. Notice also that the Euclidean theta term is purely imaginary as
it should be. With these notations we parametrise the one plaquette action as
\begin{equation}
 s(U) = - \cosh \! \epsilon \, s_w(U) + \sinh \! \epsilon \, s_\theta (U),
\end{equation}
where $\epsilon$ is a real parameter that determines the relative weight
of the two terms. This parametrisation of the action was chosen for
convenience. If we wanted the relative weight of the theta term to be the
larger one, the convenient choice would be to exhange the cosh and the
sinh. The calculation would go in essentially the same way.

In the reaminder of this section we explicitly calculate the integrals
(\ref{eq:charcoeff}) with this action in both the U($N$) and the SU($N$)
case.

The integrand of (\ref{eq:charcoeff}) is a conjugacy class function so we
can rewrite it in terms of the eigenvalues of $U$, $\{ \exp(i\alpha_1),
\exp(i\alpha_2),...,\exp(i\alpha_N)\} $. In terms of these variables
the action is
\begin{equation}
 s(\vec{\alpha}) =  - \sum_{j=1}^N ( \cosh \! \epsilon \,
 \cos \alpha_j - i \sinh \! \epsilon \, \sin \alpha_j )
\end{equation}
The irreducible representations of $U(N)$ can be labelled by $N$ integers
$\{ n_1 \geq n_2 \geq,...,\geq n_N \} = [n]$. According to Weyl's character
formula the character of this representation reads
\begin{equation}
 \chi_{[n]}(\vec{\alpha}) = \frac{\det_{jk} \mbox{e}^{i \alpha_j
 (n_k+N-k)}} { \Delta(\vec{\alpha}) },
\end{equation}
where
\begin{equation}
 \Delta(\vec{\alpha}) = \det_{jk} \mbox{e}^{i \alpha_j (N-k)}.
      \label{eq:delta}
\end{equation}
Finally the invariant measure on $U(N)$ is given by
\begin{equation}
 dU \propto \prod_{m=1}^N \frac{d \alpha_m}{2 \pi} \;
 | \Delta(\vec{\alpha}) |^2,
     \label{eq:Haar}
\end{equation}
up to a factor independent of the $\alpha$'s. Putting this all together
and introducing
\begin{equation}
 f(\alpha) = \frac{N}{g^2} (\cosh \! \epsilon \, \cos \alpha -
                          i \sinh \! \epsilon \, \sin \alpha )
\end{equation}
as a shorthand notation we obtain
\begin{eqnarray}
 d_{[n]} \Lambda_{[n]}(N/g^2) =  \mbox{const.\ } \prod_{l=1}^N
 \int_{-\pi}^{\pi} \frac{d \alpha_l}{2 \pi} \mbox{e}^{f(\alpha_l)}
 \Delta(\vec{\alpha})  \; (\det_{jk} \mbox{e}^{i \alpha_j (n_k+N-k)} )^*
 \hspace{2cm}
                                 \nonumber \\[2mm]
  = \mbox{const.} \prod_{l=1}^N \int_{-\pi}^{\pi} \frac{d \alpha}{2 \pi}
 \mbox{e}^{f(\alpha_l)} \left[ \sum_{\sigma \in S_N} (-1)^\sigma \prod_{j=1}^N
 \mbox{e}^{i \alpha_j (N-\sigma_j)}  \right] \hspace{5cm}
                                 \nonumber \\[2mm]
 \times \left[ \sum_{\eta \in S_N}
 (-1)^\eta \prod_{m=1}^N \mbox{e}^{-i \alpha_m (n_{\eta(m)}+N-\eta(m))}
 \right]                         \nonumber \\[2mm]
 = \mbox{const.} \sum_{\mu \in S_N} (-1)^\mu \sum_{\sigma \in S_N}
 \prod_{l=1}^N \int_{-\pi}^\pi \frac{d \alpha_l}{2 \pi} \exp \left[
 f(\alpha_l) - i\alpha_l \; ( \sigma(l) +n_{\mu \sigma(l)} -\mu \sigma(l))
 \right],
\end{eqnarray}
where the summations are on all the permutations of $N$ elements. The
terms of the last sum on $\sigma$ are easily seen to be independent of
$\sigma$ thus
\begin{equation}
 d_{[n]} \Lambda_{[n]} (N/g^2) = \mbox{const.} \; N! \; \det_{jk} \left[
 \int_{-\pi}^{\pi} \frac{d \alpha}{2 \pi} \, \mbox{e}^{f(\alpha) -i\alpha
 (n_j+k-j)} \right]
      \label{eq:lambda}
\end{equation}
Now all that remains to be done is to calculate the Fourier expansion
of e$^{f(\alpha)}$
\begin{eqnarray}
 \int_{-\pi}^{\pi} \frac{d \alpha}{2 \pi} \; \exp \left[ \frac{N}{g^2}
 ( \sinh \! \epsilon \, \cos \alpha - i \cosh \! \epsilon \, \sin \alpha)
 -im \alpha \right]   \nonumber \\[2mm]
 = \int_{-\pi}^{\pi} \frac{d \alpha}{2 \pi} \, \exp \left[ \frac{N}{g^2}
 \cos(\alpha + i \epsilon) -im \alpha \right] =
   \mbox{e}^{-m \epsilon} I_m(N/g^2).
\end{eqnarray}
Here $I_m$ is the modified Bessel function of order $m$. Substituting this
into (\ref{eq:lambda}) we obtain the final expression for the U($N$)
character expansion coefficients
\begin{equation}
 \Lambda_{[n]}^{U(N)}(g) = \frac{1}{d_{[n]}} \det_{jk} \left(
 \mbox{e}^{-\epsilon (n_j-j+k)} \, I_{n_j-j+k}(N/g^2) \right)
     \label{eq:lambdaUN}
\end{equation}

In the $SU(N)$ case there are two differences compared to U($N$). At first
the irreducible representations of $SU(N)$ are labelled by non-negative
integers satisfying $n_1 \geq n_2 \geq,...\geq n_N=0$, which does not
change anything in the previous calculation. Secondly for SU($N$) there is
an additional $\sum_{k=1}^N \alpha_k =0$ constraint on the eigenvalues
that must be enforced by a delta function in the measure. If this delta
function is written as
\begin{equation}
 \delta(\sum_{k=1}^N \alpha_k) = \sum_{m=-\infty}^{\infty} \mbox{e}^{
 im \sum_{k=1}^N \alpha_k},
\end{equation}
the $U(N)$ calculation goes through without any modification giving the
final result
\begin{equation}
 \Lambda_{[n]}^{SU(N)}(N/g^2) = \frac{1}{d_{[n]}} \sum_{m=-\infty}^{\infty}
 \det_{jk} \left[ \mbox{e}^{-\epsilon (n_j-j+k+m)} \, I_{n_j-j+k+m}(N/g^2).
 \right]
     \label{eq:lambdaSUN}
\end{equation}
 In the $\epsilon \rightarrow 0$ limit the action is pure Wilson and
the expressions we obtained give the correct character expansion of the
Wilson action. Had we chosen to give a bigger relative weight to the
theta term by exchanging the cosh and the sinh,
we would have gotten the same result but with Bessel functions instead
of modified Bessel functions.

\section{The continuum limit}

In this section we calculate the continuum limit of the character expansion
coefficients to leading (nontrivial) order in $g \rightarrow 0$. In 2d
gauge theories the convergence radius of the strong coupling expansion
is infinite and it makes sense to calculate the weak coupling limit of the
terms in a strong coupling expansion.

As $g \rightarrow 0$ we want the coupling of the theta term to approach
its continuum value $\theta/2 \pi$. Accordingly $\epsilon$ has to be scaled
in this limit as
\begin{equation}
 \epsilon \propto \frac{\theta}{2 \pi} \, \frac{g^2}{N}.
     \label{eq:epsilon}
\end{equation}
This can also be seen from dimensional analysis. For our purpose
the only ``interesting'' part of the $\Lambda_{[n]}$'s is the one
that depends on the representation label. From now on we drop
representation independent factors and whenever we do so the ``$\approx$''
sign will be used. Later on we shall comment on the information that
might have been lost due to these omittions.

Substituting the asymptotic representation of the modified Bessel functions
\begin{equation}
 I_n(z) \longrightarrow \frac{e^z}{\sqrt{2 \pi z}} \;
 \mbox{e}^{-\frac{n^2}{2z}} \hspace{1cm} \mbox{as} \hspace{5mm} z
 \rightarrow \infty
\end{equation}
and (\ref{eq:epsilon}) for $\epsilon$ into (\ref{eq:lambdaUN}),
for small $g$ we obtain
\begin{eqnarray}
 \Lambda_{[n]}(g,\theta) & \approx & \frac{1}{d_{[n]}} \det_{jk} \,
 \mbox{e}^{ -\frac{g^2}{2N} \left[ (n_j-j+k)^2 - \frac{\theta}{\pi}
 (n_j-j+k) \right]}
                              \nonumber \\
 & \approx & \frac{1}{d_{[n]}} \exp \left( -\frac{g^2}{2N} \sum_{j=1}^N
 n_j (n_j -2j+2- \frac{\theta}{\pi}) \right) \, \det_{jk} \mbox{e}^{
 -\frac{g^2}{N} (n_j-j)(k-1)}
      \label{eq:det}
\end{eqnarray}
In the last step the determinant was expanded as a sum over permutations,
a factor, independent of the permutations was pulled out and the remainder
was rewritten as a determinant. The last factor in (\ref{eq:det}) is a
Van der Monde determinant which can be expressed as
\begin{equation}
 \det_{jk} \mbox{e}^{-\frac{g^2}{N} (n_j-j)(k-1)} =
 \prod_{j<k} \left( \mbox{e}^{-\frac{g^2}{N} (n_j-j)} -
 \mbox{e}^{-\frac{g^2}{N} (n_k-k)} \right).
\end{equation}
The last ingredient that we have not given yet is the dimension of the
representation $[n]$,
\begin{equation}
 d_{[n]} \approx \prod_{j<k} \left( n_k-k -n_j+j \right).
\end{equation}
Making use of this and the asymptotic expansion
\begin{equation}
 \ln \left( \frac{ e^{-ax} - e^{-bx}}{b-a} \right) \longrightarrow
 \ln x  \; - \frac{a+b}{2} x + ... \hspace{1cm} \mbox{as}
 \hspace{5mm} x \rightarrow 0
\end{equation}
one then obtains
\begin{equation}
 \frac{1}{d_{[n]}} \det_{jk} \mbox{e}^{-\frac{g^2}{N} (n_j-j)(k-1)}
 \approx \exp \left( -\frac{g^2}{N} (N-1) \sum_{j=1}^N n_j \right)
\end{equation}
This completes the calculation of $\Lambda_{[n]}^{U(N)}$ in the continuum
limit, giving
\begin{equation}
 \Lambda_{[n]}^{U(N)} (g,\theta) \approx \exp \left( -\frac{g^2}{2N}
 \sum_{j=1}^N n_j(n_j-2j+N+1 - \frac{\theta}{\pi}) \right).
     \label{eq:finalUN}
\end{equation}

In the exponent for $\theta=0$ we recover the eigenvalue of the quadratic
Casimir which gives the heat kernel action.

The previous calculation goes through in the SU($N$) case without any
modification, except that we have to do a little bit more work to
evaluate the infinite sum in (\ref{eq:lambdaSUN}).
\begin{eqnarray}
 \Lambda_{[n]}^{SU(N)} (g,\theta) & \approx & \sum_{m=-\infty}^{\infty}
 \exp -\frac{g^2}{2N} \sum_{j=1}^N (n_j+m)(n_j+m-2j+N+1-
 \frac{\theta}{\pi} )
                          \nonumber \\[2mm]
 & = & \Lambda_{[n]}^{U(N)} \; \sum_{m=-\infty}^{\infty}
 \exp -\frac{g^2}{2N} \left( Nm^2 +2m \sum_{j=1}^N (n_j
 -\frac{\theta}{2 \pi} ) \right).
\end{eqnarray}
In the $g \rightarrow 0$ limit this sum can be explicitly evaluated by
approximating it with a Gaussian integral. After calculating this integral
we obtain
\begin{eqnarray}
 \Lambda_{[n]}^{SU(N)} & \approx & \exp - \frac{g^2}{2N} \left[
 \sum_{j=1}^N n_j(n_j-2j+N+1- \frac{\theta}{\pi})
 - \frac{1}{N} \left( \sum_{j=1}^N n_j + \frac{N \theta^2}{4 \pi^2}
  \right)^2  \right]               \nonumber \\[2mm]
 & = & \exp -\frac{g^2}{2N} \left[ \sum_{j=1}^N n_j(n_j-2j+N+1)
 -\frac{\left( \sum_{j=1}^N n_j \right)^2}{N} - N\frac{\theta^2}{4 \pi^2}
 \right]
\end{eqnarray}
Again the $\theta=0$ case agrees with the character expansion of the heat
kernel action.

Notice that while in the U($N$) case $\theta$ couples to the representations
through their diagonal U(1) charge $\sum_{j=1}^N n_j$, in the SU($N$)
case $\theta$ appears only in a representation independent term that
can as well be dropped. This is exactly what we expected. Had we formulated
the model in the continuum from the outset, the theta term would have been
identically zero for SU($N$). The reason for this is that the continuum
topological term is proportional to the integral of the trace of the
field strength and the trace vanishes for SU($N$).

Let us now take a closer look at the U($N$) case. In the previous section
we made a remark that on non-orientable surfaces only self conjugate
representations contribute to the partition sum. In terms of the
$n_j$'s the condition for a representation to be self conjugate is
\begin{equation}
 n_j = -n_{N-j+1} \hspace{1cm} \forall j: \hspace{3mm} 1 \leq j \leq N
\end{equation}
It follows that for self conjugate representations $\sum_{j=1}^N n_j =0$
and $\theta$ cannot be seen in the continuum at all. Recall that in this
case the topological charge must be identically zero because the
second cohomology group of any compact closed non-orientable 2d manifold
is trivial. Thus the integral of any closed 2-form and in
particular that of the curvature must be zero. Indeed we do not expect
the theta term to influence the continuum physics.

Finally let us deal with the only case in which there is a nontrivial
topological term in the continuum, i.e.\ the gauge group U($N$) on a
compact closed orientable surface. In this case the configurations of
the system are labelled by the irreducible representations of U($N$),
there is no additional constraint. The action corresponding to the
representation $[n]$ is given by
\begin{equation}
 \frac{Ag^2}{2N} \sum_{j=1}^N n_j(n_j-2j+N+1 - \frac{\theta}{\pi} )
 + (2g-2) \ln d_{[n]},
\end{equation}
where $A$ is the number of plaquettes forming the surface. In the
thermodynamic limit when $A \rightarrow \infty$ the last term becomes
irrelevant and the partition sum is completely dominated by the ground
state, the representation that minimises
\begin{equation}
 s_\theta [n] = \sum_{j=1}^N n_j(n_j-2j+N+1 -\frac{\theta}{\pi})
\end{equation}
If $\theta$ is varied, the ground state might change at some point(s)
due to level crossing, which causes phase transitions. This phenomenon
has already been observed in a variety of similar systems; 2d $U(1)$
gauge theory with a theta term \cite{Wiese,Wheater}, the analogous
1d spin model \cite{Asorey} and in 2d $CP(N)$ models at strong coupling
\cite{Seiberg}.

Let us now explore this possibility in the present model. For $\theta=0$
the action is
\begin{equation}
 s_0[n] = \sum_{j=1}^N n_j(n_j-2j+N+1) = \sum_{j=1}^N n_j^2 +
 \sum_{j=1}^N n_j(N+1-2j) \geq \sum_{j=1}^N n_j^2
     \label{eq:ineq}
\end{equation}
The inequality can be easily verified by pairing the terms in the previous
sum as
\begin{equation}
 \sum_{k=1}^{N/2} \left[ n_k(N+1-2k) - n_{N-k+1}(N+1-2k) \right] \geq 0
\end{equation}
and using the inequalities between the $n_k$'s. If $N$ is odd then
$k$ goes only to $\frac{N-1}{2}$  and the remaining term vanishes.
It is clear from (\ref{eq:ineq}) that at $\theta=0$ the ground state is
given by the trivial representation $n_1=...=n_N=0$. This is not surprising
since it is well known that in the strong coupling expansion of the
Wilson action the trivial representation gives the largest contribution
for any value of the gauge coupling \cite{Drouffe}.

Also from (\ref{eq:ineq}) we can see that
\begin{equation}
 s_\theta[n] \geq \sum_{j=1}^N n_j(n_j-\frac{\theta}{\pi}) =
 \sum_{j=1}^N \left( n_j -\frac{\theta}{2 \pi} \right)^2 + \mbox{const.}
\end{equation}
which shows that by increasing $\theta$ starting from zero we do not encounter
any phase transition up to $\theta=\pi$. At this point however there is a
phase transition; above $\pi$ the representation $n_1=....=n_N=1$ becomes
the ground state. Due to the periodicity in $\theta$ the same type of
phase transition occurs whenever $\theta$ becomes an odd integer times
$\pi$. The free energy density around the $\theta=\pi$ phase transition is
given by
\begin{eqnarray}
 f(\theta)=0 & 0 \leq \theta \leq \pi \nonumber \\
 f(\theta)= g^2 \left( 1 - \frac{\theta}{\pi} \right) & \theta \geq \pi
     \label{eq:free_energy}
\end{eqnarray}
We have to note that (\ref{eq:free_energy}) gives only the part of the
free energy that depends on the representation explicitly. The reason for this
is that throughout the calculation we kept only the representation
dependent terms of the free energy. Any representation independent but theta
dependent neglected terms, however, should be analytic at the phase transition.
Thus in the SU($N$) case, for example, we eventually would have needed to
include an extra $\theta$-dependent renormalisation counter-term in
the free energy to make it $\theta$-independent, as we expect it to be in the
continuum.

\section{The fermionic picture}

It is well known that 2d U($N$) QCD is esentially equivalent to a system of
$N$ fermions living on a circle \cite{Minahan}. It turns out that the result
we obtained in the previous section has a nice interpretation in this language.
The theta term produces an external magnetic field proportional to $\theta$
and the phase transition we encountered at $\theta=\pi$ corresponds to an
abrupt change of the ground state of the fermion system. This correspondance
is possibly not very surprising since the theta term and the external
magnetic field represent the same ambiguity in the quantisation of the
two equivalent systems. This ambiguity is due to the non-simply connected
nature of the classical configuration space.

In this section we briefly summarise the fermion interpretation of 2d
QCD from our point of view and discuss the relevance of the theta term.
For a discussion of the fermionic picture starting from a continuum
Hamiltonian formulation we refer the reader to \cite{Minahan}.

In section 2 we obtained the partition function of 2d QCD using the Lagrangian
description. Now we want to rewrite this in terms of the Hamiltonian in
order to identify the (Euclidean) quantum mechanical system to which this
partition function corresponds. In the Hamiltonian language the two
coordinates on the surface correspond to time and one space direction. In order
to simplify matters we first choose the simplest nontrivial surface, the
cylinder. We take the compactified dimension to be ``space'' and time
is measured along the axis of the cylinder. Using the gluing procedure
described in section 2 and the form of the character expansion coefficients
obtained in section 3 we can construct the partition function on a cylinder
with arbitrarily chosen holonomies $U_1,U_2 \in U(N)$ on the boundary
\begin{equation}
 Z_{LT}(g,U_1,U_2) = \sum_{[n]} \mbox{e}^{-\frac{LTg^2}{2N} s_\theta [n]}
 \; \chi_{[n]}(U_1) \chi_{[n]}^*(U_2),
     \label{eq:Zcylinder}
\end{equation}
where $L$ and $T$ are the size of the cylinder in the space and time direction
respectively. In the language of quantum mechanics the partition function of
a system with fixed configurations at $t=0$ and $t=T$ is the (Euclidean)
propagator
\begin{equation}
 \langle q_2 T|q_10 \rangle = \sum_j \langle q_2|j \rangle \; \;
 \mbox{e}^{-E_jT} \; \;  \langle j|q_1 \rangle,
     \label{eq:propagator}
\end{equation}
where the $|j \rangle$'s form a complete set of eigenstates of the Hamiltonian
and the $E_j$'s are the corresponding energies.
Comparing (\ref{eq:Zcylinder}) and
(\ref{eq:propagator}) we conclude that in 2d QCD the energy eigenstates are
labelled by the irreducible representations and the corresponding
eigenfunctions are the characters.

In this formulation the wave function is a function on the gauge group
which might suggest that the classical configuration space is the group
itself. This is however not quite right because the characters do not form
a complete set in the space L$^2(U(N))$, they span only the subspace of
conjugation invariant functions on U($N$). If we want to interpret the
partition function of the cylinder as the propagator of a quantum mechanical
system, the wave functions have to be restricted to be invariant under
conjugation. This restriction looks more natural from the gauge theory
point of view. At any fixed time $t$ the only (independent) gauge invariant
observable is the conjugacy class of the holonomy around the ``space'' circle
at $t$. It follows that the classical configuration space is not the whole
gauge group but only the set of its conjugacy classes. The most convenient
parametrisation of this set is given by the phases of the eigenvalues of the
group elements as described in section 2. In this way the classical
configuration space in the U($N$) case can be identified with the
$N$-torus $T^N$ (a maximal torus in U($N$)) factorised by the
permutation group of $N$ elements. This
factorisation is essential because all the permutations of a given set
of eigenvalues correspond to the same conjugacy class. The classical
configuration space is thus equivalent to that of a system of $N$
indistinguishable particles on a circle and the wave functions are symmetric
functions on $T^N$.

There is however one more complication due to the nontrivial nature of the
integration measure defining the inner product on the Hilbert space. Recall
that the invariant measure in terms of the above given coordinate system
is given by (\ref{eq:Haar}). The inner product is
\begin{equation}
 \langle \Psi | \Phi \rangle = \prod_{m=1}^N \int_{-\pi}^{\pi}
 \frac{d \alpha_j}{2 \pi} \Psi^*(\vec{\alpha}) \; \Delta^*(\vec{\alpha}) \;
 \Delta(\vec{\alpha}) \; \Psi(\vec{\alpha}),
\end{equation}
where $\Delta(\vec{\alpha})$ is the determinant (\ref{eq:delta}).
It is now seen that by including the determinant $\Delta$ in the
wave function the inner product is written in terms of the ``usual''
measure on $T^N$. Since the original wave functions $\Psi$ were symmetric
the redefined wave functions $\Psi \Delta$ will be antisymmetric. This
shows that the particles we are dealing with are fermions.

To see that it is actually the whole Hilbert space of $N$ fermions that
enters the expression for the propagator we have to look at the irreducible
characters after the above redefinition
\begin{equation}
 \chi_{[n]}(\vec{\alpha}) \;  \Delta(\vec{\alpha}) =
 \det_{jk} \mbox{e}^{i \alpha_j (n_k+N-k)} = \mbox{e}^{i \frac{N-1}{2}
 \sum_{j=1}^N \alpha_j }  \; \det_{jk} \mbox{e}^{i \alpha_j (n_k +
 \frac{N+1}{2} - k)}
     \label{eq:Slater}
\end{equation}
and verify that they span the whole Hilbert space. (\ref{eq:Slater}) is
essentially a Slater determinant composed of one-particle momentum eigenstates
e$^{i\alpha_j p_k}$ with momenta
\begin{equation}
 p_k = n_k + \frac{N+1}{2} -k.
\end{equation}
For N odd it is easily checked that the irreducible representations are
indeed in one-to-one correspondance with the states in the $N$-fermion
Fock space on a unit circle. If $N$ is even the allowed momenta are half
integers which give antiperiodic boundary conditions for the $N$-fermion
wave function.

The wave function (\ref{eq:Slater}) describes an energy eigenstate with
energy
\begin{equation}
 E[p] = \frac{Lg^2}{2N} s_\theta[n] = \frac{Lg^2}{2N} \left[
 \sum_{j=1}^N \left( p_j - \frac{\theta}{2 \pi} \right) ^2 - \sum_{j=1}^N
 \left( \frac{N+1}{2} - j - \frac{\theta}{2 \pi} \right) ^2 \right]
\end{equation}
The first term is the energy of $N$ noninteracting particles in the
presence of a magnetic field with constant $\theta/2\pi$ vector potential
along the circle on which the particles live. The second term can be
thought of as a trivial -- though $\theta$-dependent -- shift of the energy
that does not depend on the state of the system. For $\theta< \pi$ the ground
state is the usual fermionic ground state with all the states filled below
the Fermi surface. The Fermi surface in this case degenerates to the two
points $\pm p_F = \pm \frac{N-1}{2}$. In summary, we have succeded in showing
that 2d U($N$) QCD is equivalent to a system of $N$ noninteracting fermions
moving on a unit circle. The correspondance between the amplitudes in the
fermionic model and the gauge model is
\begin{equation}
 \langle \vec{\beta} \, T | \vec{\alpha} \, 0 \,
 \rangle_{\mbox{\tiny fermion}} =
 \Delta^*(\vec{\beta}) \; \; \Delta(\vec{\alpha}) \;\;
 \langle U(\vec{\beta})\,  T | U(\vec{\alpha})\, 0 \rangle_{\mbox{\tiny gauge}}
\end{equation}

In this language the partition function on the torus can be obtained by
taking periodic boundary condition in the time direction and integrating
out the holonomy around the two identified boundary pieces
\begin{equation}
 Z_{\mbox{\tiny torus}}(g) = \sum_{[p]} \mbox{e}^{-TE[p] }
     \label{eq:Ztorus}
\end{equation}
This is the partition function of the corresponding fermion gas at temperature
$1/T$. In the zero temperature limit only the ground state contributes to
(\ref{eq:Ztorus}). At $\theta=\pi$ however the ground state becomes degenerate
and above $\pi$ the new ground state will be the one with the momenta of all
the particles shifted by one unit. This is the phase transition that we
encountered in the previous section.

Equation (\ref{eq:Zcylinder}) gives the propagator in terms of momentum
eigenstates. As a final check of our results it is instructive to rewrite
the $N$-fermion propagator using position eigenstates. This can be done
by using the Poisson summation formula
\begin{equation}
 \sum_{p=-\infty}^\infty f(p) = \sum_{m=-\infty}^\infty \;\;
 \int_{-\infty}^\infty dx \, f(x) \, \mbox{e}^{2 \pi imx}
\end{equation}
The constraint $p_1>p_2...$ on the momenta can be dropped at the expense of
a factor $N!$ and we also omit the representation-independent term in
$E[p]$. Using the notation $\Delta_{[p]}$ for the Slater determinant
the fermion propagator reads
\begin{eqnarray}
   \langle \vec{\beta} \, T \, | \, \vec{\alpha} \, 0 \rangle \;\;
 \mbox{e}^{i \frac{N-1}{2} \sum_{j=1}^N (\alpha_j - \beta_j)} =
 \sum_{p_j=-\infty}^\infty \; \Delta^*_{[p]}(\vec{\beta}) \;
 \mbox{e}^{-\frac{LTg^2}{2N} \sum_{j=1}^N (p_j-\frac{\theta}{2 \pi})^2} \;
 \Delta_{[p]}(\vec{\alpha})     \nonumber  \\[2mm]
  = \frac{1}{N!} \; \sum_{m_j=-\infty}^\infty \; \sum_{\sigma,\eta \in S_N}
 (-1)^{\sigma \eta} \; \prod_{j=1}^N  \left[ \int_{-\infty}^\infty dy_j \,
 \mbox{e}^{-\frac{TLg^2}{2N} y_j^2 + iy_j (2 \pi m_j + \alpha_{\sigma(j)}
 -\beta_{\eta(j)} + \frac{iTLg^2 \theta}{2N \pi})}   \right]
                                \nonumber \\[2mm]
  = \frac{1}{N!} \; \left( \frac{2N \pi}{TLg^2} \right)^{\frac{N}{2}} \;
 \sum_{m_j=-\infty}^\infty \; \mbox{e}^{-\frac{i\theta}{2 \pi} \sum_{j=1}^N
 (2\pi m_j+\alpha_j-\beta_j)} \hspace{5cm}          \nonumber \\[2mm]
  \times \sum_{\sigma,\eta \in S_N} (-1)^{\sigma \eta} \exp
 \left[ -\frac{N}{2TLg^2} \sum_{j=1}^N (2\pi m_j+ \alpha_{\sigma(j)}
 - \beta_{\eta(j)} )^2 \right]
                          \label{eq:prop}
\end{eqnarray}
This is the familiar form of the Euclidean propagator for a system of $N$
fermions on a circle. Each term in the sum over $m$-configurations
corresponds to the contribution of paths with a particular set of
winding numbers $[m_1, m_2,...,m_N]$. In complete agreement with our
expectation the theta term gives an extra phase factor to each winding
sector.

Recall that if the configuration space of the classical system is not
simlply connected quantisation becomes ambiguous. Paths belonging to
different homotopy classes can get extra relative phase factors that
belong to a representation of the first homotopy group of the classical
configuration space. The ambiguity arises because there is no a priori
criterium to select a particular representation \cite{Schulman}. The
$\theta$-dependent phase factor in (\ref{eq:prop}) represents exactly
this quantisation ambiguity, the representations of the homotopy group
being labelled by $\theta$. This concludes our discussion of the fermion
gas interpretation of 2d U($N$) QCD.

\section{Conclusions}

In this paper we have shown a way of incorporating a theta term in
two dimensional U($N$) and SU($N$) gauge theories. Starting from a
lattice formulation we derived the continuum limit of the action
resulting in a generalisation of the heat kernel action in the presence
of a topological term. Our choice of the lattice theta term was the most
naive, the only requirement it satisfied was having the topological term
(first Chern number) as its naive continuum limit. In this way the
lattice ``topological charge'' was not necessarily an integer and it was
not necessarily zero in the topologically trivial cases. In spite of this,
in the continuum limit we have been able to recover the topological
properties of the system. In the topologically trivial cases when the
topological charge was expected to vanish identically (SU($N$) gauge group
on any surface and U($N$) gauge group on non-orientable surfaces) the
only effect of the theta term remaining in the continuum limit was a
trivial constatant shift of the ground state energy. On the other hand in the
topologically nontrivial case (U($N$) gauge group on orientable surfaces)
the theta term remained to be coupled to the dynamics in the continuum,
giving rise to phase transitions at $\theta_c = \pm \pi, \pm 3 \pi, \pm 5 \pi
...$ The same type of phase transition has been observed in CP($N$) models
at strong coupling \cite{Seiberg}. It is not very surprising that in the
2d gauge system even in the weak coupling limit we obtain an essentially
strongly coupled system. This is due to the fact that in 2d QCD the
strong coupling expansion has an infinite radius of convergence and we
expect the strong coupling behaviour to persist down to any nonzero value
of the coupling.

Recently, based on Monte Carlo calculations, Schierholz argued that
in the two dimensional CP(3) model and in four dimensional SU(3) QCD
this phase transition also takes place but the critical value of $\theta$
becomes small for weak coupling, eventually reaching zero in the
continuum limit \cite{Schierholz}.
In this way $\theta$ is forced to be zero if confinement is to be maintained
and this gives a solution to the strong CP problem. From
the argument given at the end of the previous paragraph it is clear that our
results for 2d QCD give no direct hint concerning the phase transition in 2d
CP($N$) or 4d QCD. Such a weak coupling transition must be directly related to
the mechanisms responsible for maintaining confinement at arbitrarily weak
coupling in four dimensions. We hope to address this question elsewhere.

In the previous section we interpreted the continuum theta term in the
fermionic formulation of 2d U($N$) QCD. In particular we have shown that
in the fermion language the theta term corresponds to an external magnetic
field imposed on the fermion gas. Very recently the connection between
the gauge model an the fermion gas has been exploited to obtain some
new results concerning the Calogero-Sutherland model \cite{Minahan2}.
Our results might also find some application in this direction.

\section*{Acknowledgements}

One of the authors (T.G.K.) is grateful to J.\ Wheater for introducing him
to this problem. This research was supported partly by NSF Grant
\# PHY-92-18990 and OTKA (Hungarian Research Fund) \# F14276.

\end{document}